\documentclass{hep99}
\usepackage{epsfig}
\usepackage{amssymb}
\begin{document}

\title{Constraints from the muon anomalous magnetic moment and $b \rightarrow s \gamma$ on gauge-mediated supersymmetry-breaking models}
\author{K.T. Mahanthappa and Sechul Oh}
\address{Department of Physics, University of Colorado, Boulder, CO 80309, USA 
\\[3pt]
E-mails: {\tt ktm@verb.colorado.edu, ohs@spot.colorado.edu}}

\abstract{We present a combined study of the muon anomalous magnetic moment, $a_{\mu} \equiv (g-2)_{\mu} /2$, and $b \rightarrow s \gamma$ decay in the minimal supersymmetric standard model with gauge-mediated supersymmetry-breaking. 
Combining new experimental data on $a_{\mu}$ and the branching ratio for $b \rightarrow s \gamma$, useful limits on the parameter space of these models are derived. Bounds on supersymmetric particle masses as a function of $\tan \beta$ are also presented. } 

\maketitle

\section{Introduction}
The gauge-mediated supersymmetry breaking (GMSB) models have been of special interest, because they have attractive features of natural suppression of the supersymmetry (SUSY) contributions to flavor-changing neutral currents at low energies and prediction of the supersymmetric particle mass spectrum in terms of few parameters.  

The decay process $b \rightarrow s \gamma$ does not occur at the tree level, and at one-loop level it occurs at a small rate but enough to be sensitive to effects of new physics.
The CLEO collaboration has recently reported the branching ratio for the decay $b \rightarrow s \gamma$ \cite{2} : $2.0 \times 10^{-4} < {\rm BR} (b \rightarrow s\gamma) < 4.5 \times 10^{-4}$ at 95 $\%$ C.L.  The anomalous magnetic moment of muon, $a_{\mu} \equiv (g-2)_{\mu} /2$, is also sensitive to new physics effects and can be used to constrain SUSY models \cite{maoh,7b}, on account of the great accuracy of both experimental and the standard model (SM) theoretical values of $a_{\mu}$.  The present experimental value of $a_{\mu}$ \cite{3} is $a^{\rm exp}_{\mu} = 11659230(84) \times 10^{-10}$, while the theoretical prediction for $a_{\mu}$ in  the context of the SM  is $a^{\rm SM}_{\mu} = 11659162(6.5) \times 10^{-10}$ \cite{maoh}.     
  
In this work, we obtain combined constraints due to both $b \rightarrow s \gamma$ decay and $a_{\mu}$ in the minimal supersymmetric SM (MSSM) with GMSB.  Even though there exist the previous works which studied either $b \rightarrow s \gamma$ \cite{maoh,6a} or $a_{\mu}$ \cite{7b} in the GMSB models, our work extends the previous ones in the sense that we investigate both $b \rightarrow s \gamma$ and $a_{\mu}$ together with the inclusion of the supersymmetric one-loop correction to the mass of $b$ quark, $m_b$, which has considerable effects in large $\tan \beta$ region as we see below.  Furthermore, in this combined study, we explicitly show that, with the presently available experimental data, constraints from the decay $b \rightarrow s \gamma$ are more stringent than those from $a_{\mu}$ in broad region of the parameter space. 

\section{The model}
In the GMSB models messenger fields transmit SUSY breaking to the fields of visible sector via loop diagrams involving SU(3)$_C \times$SU(2)$_L \times$U(1)$_Y$ gauge interactions.  
 
The radiatively generated soft SUSY-breaking masses of gauginos and scalars at messenger scale $M$ are given in terms of $\Lambda= F/M$ ( $\sqrt{F}$ is the original SUSY-breaking scale ), the SM gauge couplings $\alpha_i$ $(i=1,2,3)$ and the effective number of messenger fields $n$ $[$ $n =n_5 +3 n_{10}$, where $n_5$ and $n_{10}$ denote the number of (${\bf 5} +\bar {\bf 5}$) and (${\bf 10} +\overline{{\bf 10}}$) pairs, respectively $]$.  
It is known that for messenger fields in complete SU(5) representation, at most 
four (${\bf 5} +\bar {\bf 5}$) pairs, or one (${\bf 5} +\bar {\bf 5}$) and one (${\bf 10} +\overline{{\bf 10}}$) pair are allowed to ensure that the gauge couplings remain perturbative up to the GUT scale.

\section{The analysis}
We require that electroweak symmetry be radiatively broken.  
The parameter $\Lambda$ is taken to be around 100 TeV to ensure that the sparticle masses are of the order of the weak scale.  The case $M = \Lambda$ is excluded since it produces a massless scalar in the messenger sector.  The upper bound on the gravitino mass of about $10^4$ eV restricts $M / \Lambda < 10^4$.        
In running the renormalization group equations, we include the one-loop correction to the running bottom quark mass, $\Delta m_b$, which involves the contributions coming from gluino-sbottom loop diagram and chargino-stop loop diagram.  

In $b \rightarrow s\gamma$ decay, the contributions to the total decay amplitude are coming from the $W$ loop diagram, charged Higgs boson loop diagram, neutralino loop diagram, and gluino loop diagram.  It has been pointed out that the neutralino and gluino contributions to the amplitude are less than 1 $\%$ in the whole range of parameter space \cite{6a}.  The charged Higgs boson loop contribution adds constructively to the $W$ loop contribution, while the chargino loop contribution can be constructive or destructive to the $W$ loop contribution, but is generally much smaller than the charged Higgs boson loop contribution.  

The supersymmetric contributions, $\delta a^{\rm SUSY}_{\mu}$, to the muon anomalous magnetic moment are essentially coming from neutralino-smuon loop diagram and chargino-sneutrino loop diagram.  
The bound on the supersymmetric contributions to $a_{\mu}$ is given by $-71 \times 10^{-10} < \delta a^{\rm SUSY}_{\mu} < 207 \times 10^{-10}$ at 90 $\%$ C.L.  This bound is obtained by the difference between experimental value and theoretical prediction of $a_{\mu}$.  
The new E821 experiment at Brookhaven is expected to improve the experimental determination of $a_{\mu}$ to the level of $4 \times 10^{-10}$ \cite{17}. 
The electroweak contribution to $a_{\mu}$ in the SM up to two-loops is $a^{\rm EW}_{\mu} = 15.1(0.4) \times 10^{-10}$.  Any deviation from this value in the new E821 experiment could be attributed to SUSY as its contribution could be as large or larger than this value \cite{maoh}.  

We use our calculated mass spectrum and couplings to calculate the rate for $b \rightarrow s\gamma$ and $\delta a^{\rm SUSY}_{\mu}$. The results depend on physical variables $\tan \beta$, $|\mu|$, sign$(\mu)$, $M/ \Lambda$, and $n$.  
Our results for both the branching ratio for $b \rightarrow s\gamma$ and $\delta a^{\rm SUSY}_{\mu}$ are presented as a function of the weak gaugino mass $M_2$,  $|\mu|$, for fixed values of $\tan \beta$, $n$ and sign$(\mu)$. $M_2$ is directly related to $\Lambda$.  Then the bounds on the branching ratio for $b \rightarrow s\gamma$ and $\delta a^{\rm SUSY}_{\mu}$ are translated into the bounds on values of $M_2$ and $|\mu|$ in the $|\mu| - M_2$ plane for fixed values of $\tan \beta$, $n$ and sign$(\mu)$.  
Bounds on other sparticle masses can be easily deduced from a bound on $M_2$.  

In Figs. 1$-$3, we display the bounds obtained from the branching ratio for $b \rightarrow s\gamma$ and $\delta a^{\rm SUSY}_{\mu}$ in the $|\mu| - M_2$ plane for $n=1$ and either sign of $\mu$, for each of the values of $\tan \beta = 10$ and 60.  
Solid lines represent the bounds from the branching ratio for $b \rightarrow s\gamma$ and dot-dashed lines describe the bounds from $\delta a^{\rm SUSY}_{\mu}$. 

\begin{figure}
\begin{center}
\epsfig{file=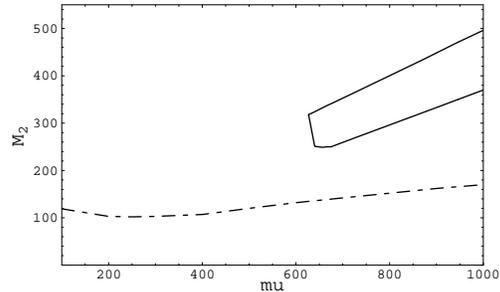, width=6.5cm}
\end{center}
\caption{Limits on the weak gaugino mass $M_2$ as a function of $|\mu|$ for $\tan \beta =10$, $\mu >0$ and $n=1$.  Units are in GeV.  Solid lines represent the bounds from the branching ratio for $b \rightarrow s\gamma$ (the region surrounded by the solid line is allowed) and dot-dashed lines represent the lower bounds from $a_{\mu}$.}
\end{figure}

Figs. 1 and 2 show the bounds on $M_2$ and $|\mu|$ for $\tan \beta =10$ and $n=1$, and for positive and negative $\mu$, respectively.  The region surrounded by the solid line is allowed by the CLEO bound, while the upper region of the dot-dashed line is allowed by the present bound on  $a_{\mu}$.  In the case of Fig. 1, the constraint from $b \rightarrow s\gamma$ decay is clearly much stronger than that from $a_{\mu}$.  We find $M_2 > 248$ GeV and $\mu > 626$ GeV.  Small values of $M_2$ lead to unacceptably large contribution to the branching ratio for $b \rightarrow s\gamma$, while large values of $\mu$ raise the problem of fine-tuning and are generally constrained by the lower bound on the stau mass.  
In Fig. 2 we see the constrains from both $b \rightarrow s\gamma$ and $a_{\mu}$ are complementary.  By combining the bounds from the both, we can obtain much stronger bound on $M_2$ and $|\mu|$; in particular, low values of $|\mu|$ which would have been allowed are excluded.  We find $M_2 > 210$ GeV and $|\mu| > 505$ GeV.  

\begin{figure}
\begin{center}
\epsfig{file=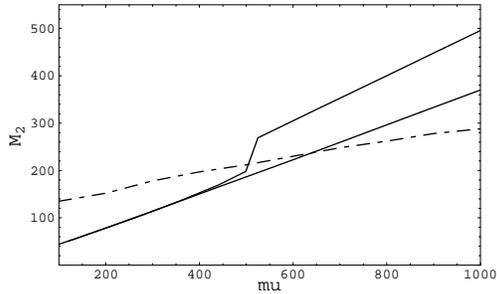, width=6.5cm}
\end{center}
\caption{The same as Fig. 1, except $\mu <0$.}
\end{figure}

In large $\tan \beta$ case, we find that the bound from either $b \rightarrow s\gamma$ or $a_{\mu}$ is more stringent than that in small $\tan \beta$ case, and most region in the $|\mu| - M_2$ plane is excluded.  For $\tan \beta = 60$ and $\mu >0$ (Fig. 3), the allowed regions from each of $b \rightarrow s\gamma$ and $a_{\mu}$ do not overlap, even though a possibility exists that they might overlap for unacceptably very large values of $\mu$.  
Thus, this case is excluded, while it would be allowed if one considered only either $b \rightarrow s\gamma$ or $a_{\mu}$ as in Refs. \cite{7b,6a}.  
For $\tan \beta \lesssim 50$ and $\mu >0$, the allowed regions from each of $b \rightarrow s\gamma$ and $a_{\mu}$ overlap allowing limited regions in the parameter space.  
For $\tan \beta = 60$ and $\mu <0$, the supersymmetric one-loop correction to bottom quark mass leads to unacceptably large value of $m_b$, unless one makes additional assumptions like $b - \tau$ Yukawa coupling unification \cite{maoh}. 
To keep our analysis in a general form in the context of the GMSB models, we adopt no further assumptions like $b - \tau$ unification.  
Thus, by inclusion of the correction $\Delta m_b$, we exclude the case of large $\tan \beta$ and $\mu <0$.  
For $n=3$ \cite{maoh}, large $\tan \beta$ region is almost ruled out due to the same reason as the case of $n=1$.  

\begin{figure}
\begin{center}
\epsfig{file=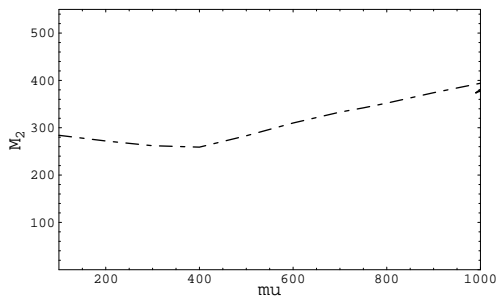, width=6.5cm}
\end{center}
\caption{The same as Fig. 1, except $\tan \beta =60$.}
\end{figure}

\begin{figure}
\begin{center}
\epsfig{file=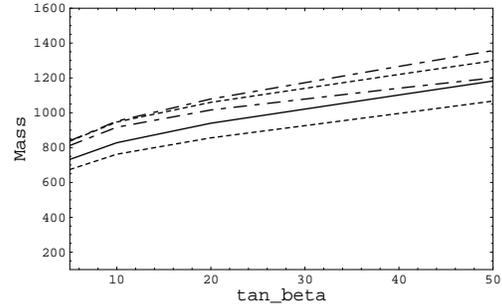, width=6.5cm}
\end{center}
\caption{Bounds on the sparticle masses (in GeV) as a function of $\tan \beta$ for $\mu >0$ and $n=1$.  The solid line represents the lower bound on the gluino mass, and the dotted and dot-dashed lines represent the lower bounds on the stop ($m_{\tilde t_1}$ and $m_{\tilde t_2}$) and sbottom ($m_{\tilde b_1}$ and $m_{\tilde b_2}$) masses, respectively.}
\end{figure}

In Fig. 4 we plot the bounds on the sparticle masses, obtained by this combined analysis of $b \rightarrow s\gamma$ and $a_{\mu}$, as a function of $\tan \beta$ for positive $\mu$ and $n=1$.  The plots are displayed for up to $\tan \beta \approx 50$, since the region corresponding to $\tan \beta \gtrsim 50$ is ruled out.  
The lower bounds on the sparticle masses increase monotonically as $\tan \beta$ does.  For $n=3$, the lower bound on each sparticle mass is higher than that for $n=1$. 

\section{Conclusion}
In our analysis, the large $\tan \beta$ region is ruled out or severely constrained, depending on the sign of $\mu$.  
By inclusion of the supersymmetric one-loop correction to $b$ quark mass, we have found that the region of large $\tan \beta$ and negative $\mu$ is physically ruled out in order to give a correct value of $m_b$, unless one makes further assumptions such as $b - \tau$ Yukawa coupling unification.  
With the present experimental data for $b \rightarrow s\gamma$ and $a_{\mu}$, constraints from the decay $b \rightarrow s \gamma$ are more stringent than those from $a_{\mu}$ in broad region of the parameter space.  However, if the Brookhaven E821 experiment approaches the expected precision of the level of $4 \times 10^{-10}$ in determination of $a_{\mu}$ in near future, constraints from $a_{\mu}$ are expected to become much more stringent than the present ones. 
We could anticipate more severe constraints on the parameter space with the future precise measurements of the branching ratio of $b \rightarrow s\gamma$ and $a_{\mu}$.  \\

This work was supported in part by the US Department of Energy Grant No. DE FG03-95ER40894.

\end{document}